\documentclass[12pt,a4]{article}
\usepackage{epic}
\usepackage{eepic}
\usepackage{epsfig}
\psfigdriver{dvips}
\begin{document}
\renewcommand{\theequation}{\thesection.\arabic{equation}}
\thispagestyle{empty}
\vspace*{-1.5cm}
\hfill {\small KL--TH 98/10} \\[8mm]

\message{reelletc.tex (Version 1.0): Befehle zur Darstellung |R  |N, Aufruf z.B. \string\bbbr}
%
%
\message{reelletc.tex (Version 1.0): Befehle zur Darstellung |R  |N, Aufruf z.B. \string\bbbr}
%
%
%
%
%
\font \smallescriptscriptfont = cmr5
\font \smallescriptfont       = cmr5 at 7pt
\font \smalletextfont         = cmr5 at 10pt
\font \tensans                = cmss10
\font \fivesans               = cmss10 at 5pt
\font \sixsans                = cmss10 at 6pt
\font \sevensans              = cmss10 at 7pt
\font \ninesans               = cmss10 at 9pt
\newfam\sansfam
\textfont\sansfam=\tensans\scriptfont\sansfam=\sevensans
\scriptscriptfont\sansfam=\fivesans
\def\sans{\fam\sansfam\tensans}
\def\bbbr{{\rm I\!R}} 
\def\bbbn{{\rm I\!N}} 
\def\bbbE{{\rm I\!E}} 
\def\bbbm{{\rm I\!M}}
\def\bbbh{{\rm I\!H}}
\def\bbbk{{\rm I\!K}}
\def\bbbd{{\rm I\!D}}
\def\bbbp{{\rm I\!P}}
\def\bbbone{{\mathchoice {\rm 1\mskip-4mu l} {\rm 1\mskip-4mu l}
{\rm 1\mskip-4.5mu l} {\rm 1\mskip-5mu l}}}
\def\bbbc{{\mathchoice {\setbox0=\hbox{$\displaystyle\rm C$}\hbox{\hbox
to0pt{\kern0.4\wd0\vrule height0.9\ht0\hss}\box0}}
{\setbox0=\hbox{$\textstyle\rm C$}\hbox{\hbox
to0pt{\kern0.4\wd0\vrule height0.9\ht0\hss}\box0}}
{\setbox0=\hbox{$\scriptstyle\rm C$}\hbox{\hbox
to0pt{\kern0.4\wd0\vrule height0.9\ht0\hss}\box0}}
{\setbox0=\hbox{$\scriptscriptstyle\rm C$}\hbox{\hbox
to0pt{\kern0.4\wd0\vrule height0.9\ht0\hss}\box0}}}}

\def\bbbe{{\mathchoice {\setbox0=\hbox{\smalletextfont e}\hbox{\raise
0.1\ht0\hbox to0pt{\kern0.4\wd0\vrule width0.3pt height0.7\ht0\hss}\box0}}
{\setbox0=\hbox{\smalletextfont e}\hbox{\raise
0.1\ht0\hbox to0pt{\kern0.4\wd0\vrule width0.3pt height0.7\ht0\hss}\box0}}
{\setbox0=\hbox{\smallescriptfont e}\hbox{\raise
0.1\ht0\hbox to0pt{\kern0.5\wd0\vrule width0.2pt height0.7\ht0\hss}\box0}}
{\setbox0=\hbox{\smallescriptscriptfont e}\hbox{\raise
0.1\ht0\hbox to0pt{\kern0.4\wd0\vrule width0.2pt height0.7\ht0\hss}\box0}}}}

\def\bbbq{{\mathchoice {\setbox0=\hbox{$\displaystyle\rm Q$}\hbox{\raise
0.15\ht0\hbox to0pt{\kern0.4\wd0\vrule height0.8\ht0\hss}\box0}}
{\setbox0=\hbox{$\textstyle\rm Q$}\hbox{\raise
0.15\ht0\hbox to0pt{\kern0.4\wd0\vrule height0.8\ht0\hss}\box0}}
{\setbox0=\hbox{$\scriptstyle\rm Q$}\hbox{\raise
0.15\ht0\hbox to0pt{\kern0.4\wd0\vrule height0.7\ht0\hss}\box0}}
{\setbox0=\hbox{$\scriptscriptstyle\rm Q$}\hbox{\raise
0.15\ht0\hbox to0pt{\kern0.4\wd0\vrule height0.7\ht0\hss}\box0}}}}

\def\bbbt{{\mathchoice {\setbox0=\hbox{$\displaystyle\rm
T$}\hbox{\hbox to0pt{\kern0.3\wd0\vrule height0.9\ht0\hss}\box0}}
{\setbox0=\hbox{$\textstyle\rm T$}\hbox{\hbox
to0pt{\kern0.3\wd0\vrule height0.9\ht0\hss}\box0}}
{\setbox0=\hbox{$\scriptstyle\rm T$}\hbox{\hbox
to0pt{\kern0.3\wd0\vrule height0.9\ht0\hss}\box0}}
{\setbox0=\hbox{$\scriptscriptstyle\rm T$}\hbox{\hbox
to0pt{\kern0.3\wd0\vrule height0.9\ht0\hss}\box0}}}}

\def\bbbs{{\mathchoice
{\setbox0=\hbox{$\displaystyle     \rm S$}\hbox{\raise0.5\ht0\hbox
to0pt{\kern0.35\wd0\vrule height0.45\ht0\hss}\hbox
to0pt{\kern0.55\wd0\vrule height0.5\ht0\hss}\box0}}
{\setbox0=\hbox{$\textstyle        \rm S$}\hbox{\raise0.5\ht0\hbox
to0pt{\kern0.35\wd0\vrule height0.45\ht0\hss}\hbox
to0pt{\kern0.55\wd0\vrule height0.5\ht0\hss}\box0}}
{\setbox0=\hbox{$\scriptstyle      \rm S$}\hbox{\raise0.5\ht0\hbox
to0pt{\kern0.35\wd0\vrule height0.45\ht0\hss}\raise0.05\ht0\hbox
to0pt{\kern0.5\wd0\vrule height0.45\ht0\hss}\box0}}
{\setbox0=\hbox{$\scriptscriptstyle\rm S$}\hbox{\raise0.5\ht0\hbox
to0pt{\kern0.4\wd0\vrule height0.45\ht0\hss}\raise0.05\ht0\hbox
to0pt{\kern0.55\wd0\vrule height0.45\ht0\hss}\box0}}}}

\def\bbbz{{\mathchoice {\hbox{$\sans\textstyle Z\kern-0.4em Z$}}
{\hbox{$\sans\textstyle Z\kern-0.4em Z$}}
{\hbox{$\sans\scriptstyle Z\kern-0.3em Z$}}
{\hbox{$\sans\scriptscriptstyle Z\kern-0.2em Z$}}}}
\setlength{\topmargin}{-1.5cm}
\setlength{\textheight}{22cm}
\begin{center}
{\large\bf Is it possible to construct exactly solvable models?}\\
\vspace{0.5cm}
{\large O. Haschke and W. R\"uhl}\\
Department of Physics, University of Kaiserslautern, P.O.Box 3049\\
67653 Kaiserslautern, Germany \\
\vspace{5cm}
\begin{abstract}
We develop a constructive method to derive exactly solvable quantum
mechanical models of rational (Calogero) and trigonometric (Sutherland)
type. This method starts from a linear algebra problem: finding eigenvectors 
of triangular finite matrices. These eigenvectors are transcribed 
into eigenfunctions of a selfadjoint Schr\"odinger operator. We prove the 
feasibility of our method by constructing an "$AG_3$ model" of trigonometric
type (the rational case was known before from Wolfes 1975). Applying a Coxeter 
group analysis we prove its equivalence with the $B_3$ model.
In order to better understand features of our construction we exhibit the 
$F_4$ rational model with our method.
\end{abstract}
\vspace{3cm}
{\it September 1998}
\end{center}
{\it Dedicated to Professor Jan Lopusza\'nski on the occasion of his 75-th birthday}
\newpage

\section{Introduction}
The completely integrable models are traditionally characterized by their 
relation with simple Lie algebras $A_n, \, B_n, \, C_n, \, D_n, \, G_2, \, F_4, \, E_6, \,
E_7, \, E_8$. This relation is the starting point of the Hamiltonian reduction method exploited by
Olshanetsky and Perelomov \cite{1}. These models possess as limiting cases the trigonometric 
(Sutherland) and rational (Calogero) models that are exactly soluble, i.e. their eigenvalues
and eigenvectors can be derived by elementary methods.

This exact solvability has been shown to follow from the fact that the Schr\"odinger operators
can, after a "gauge transformation", be rewritten as a quadratic form of Lie algebra operators.
These Lie algebra operators are represented as differential operators acting on
polynomial spaces. This program was formulated in \cite{2} and successfully applied first to 
the $A_n$ series in \cite{3}. Then it was carried over to the other sequences $B_n, \, C_n, \, 
D_n$ and $G_2$ and even to corresponding supersymmetric models \cite{4,5}.

Our aim was to turn the arguments around and to develop an algorithm which may allow us to construct new
exactly soluble models. First investigations were presented in \cite{6}. The program contains
two major and separate issues, to render a second order differential operator curvature free
and to find a first order differential operator satisfying an integrability constraint. In
this paper we present our algorithm in the following version. We start from a standard flat
Laplacian and introduce Coxeter (or Weyl) group invariants as new coordinates. If the Coxeter group 
contains a symmetric group as subgroup, these invariants are built from elementary symmetric polynomials. 
The second order differential operators obtained this way are curvature free by construction, 
and act on polynomial spaces of these Coxeter invariants that form a flag. This flag is defined 
by means of a characteristic vector ($\vec{p}$-vector).

Then we solve the integrability constraints by constructing "prepotentials" with a fixed algorithm. These 
prepotentials define the gauge transformation alluded to above which renders the differential
operator the form of a standard Schr\"odinger operator of $N$ particles in 1-dimensional space
with a potential. Each prepotential contributes an additive term to this potential with a
free (real) coupling constant. Finally the prepotentials define the ground state wave function
of the Schr\"odinger operator which originates from the trivial polynomial in the flag and
thus contains no further information. Except a possible oscillator prepotential in the translation 
invariant cases, the prepotentials are in one-to-one relation with the orbits of the Coxeter group.  

We show that all known exactly soluble models can be obtained this way (at present we have to 
make an exemption with respect to $E_6, \, E_7, \, E_8$, but this will soon be overcome).
Applying the method of constructing the Coxeter invariants of $A_2$ \cite{4} to $A_3$, we 
obtain an "$AG_3$ model". Its Coxeter diagram is that of the affine Coxeter group $\hat{B}_3$, 
which possesses the same invariants as the Coxeter group $B_3$. This leads to an explicit proof 
of the equivalence of the $AG_3$ model with the $B_3$ model. Thus a translation invariant four--particle 
model after separation of the c.m. motion is shown to be equivalent with a translation non--invariant 
three-particle model. In this paper we also discuss $F_4$ from the view point of our algorithm. 
The Schr\"odinger operator obtained (only the rational case) deviates slightly from the one given 
in \cite{1} (probably due to a simple printing error in \cite{1}).

Thus our method shifts the centre of interest from the simple Lie algebras and their homogeneous
spaces to the corresponding Weyl groups and by generalization to the Coxeter groups. On the 
other hand, the differential operators acting on polynomial spaces of Coxeter invariants define
Lie algebras of their own, but at present these algebras are only of marginal interest.

\section{The constructive program}
We are interested here in the bound state spectrum of Schr\"odinger operators. The 
whole analysis is therefore performed in real spaces. Consider a flag of polynomial 
spaces $V_N(\vec{p}), \, N \in \bbbz_{\ge}$, $\vec{p} \in \bbbn^n$
\begin{eqnarray}
& V_N(\vec{p}) = {\rm span} \left\{ z^{r_1}_1 z^{r_2}_2 ... z^{r_n}_n | r_1p_1 + r_2p_2 + ... + r_np_n \le N \right\} &\label{1} \\
&(p_i \in \bbbn)& \nonumber
\end{eqnarray}
We consider differential operators of first order
\begin{equation}
D^{(1)}_{[\vec{\alpha};a]} = z^{[\vec{\alpha}]} \frac{\partial}{\partial z_a} \label{2}
\end{equation}
($\vec{\alpha}$ a multi-exponent) \\
and of second order
\begin{equation}
D^{(2)}_{[\vec{\alpha};a,b]} = z^{[\vec{\alpha}]} \frac{\partial^2}{\partial z_a \partial z_b}
\label{3}
\end{equation}
that leave each space $V_N(\vec{p})$ invariant. If
\begin{equation}
\vec{p} = (1,1,...,1) 
\label{4}
\end{equation}
then the operators (\ref{2}) generate the full linear (inhomogeneous) group of $\bbbr_n$ 
and the operators of second order (\ref{3}) can be obtained as products from the first 
order operators, i.e. in (\ref{2})
\begin{equation}
\vec{\alpha} = e^{(c)}, \; e^{(c)}_b = \delta^c_b \quad \rm{ or } \quad \vec{\alpha} = 0
\label{5}
\end{equation}
and in (\ref{3})
\begin{equation}
\vec{\alpha} = e^{(c)} + e^{(d)} \quad \rm{ or } \quad \vec{\alpha} = e^{(c)} \quad \rm{ or } \quad \vec{\alpha} = 0
\label{6}
\end{equation}

Now we consider a candidate for a future Schr\"odinger operator
\begin{eqnarray}
D &=& - \sum_{\vec{\alpha},a,b} g_{[\vec{\alpha};a,b]} D^{(2)}_{[\vec{\alpha};a,b]} \nonumber \\
&+& \sum_{\vec{\beta},c} h_{[\vec{\beta};c]} D^{(1)}_{[\vec{\beta};c]}
\label{2.7}
\end{eqnarray}
The eigenvectors and values of $D$ in $V_N$ can be calculated easily by finite linear 
algebra methods. Let 
\begin{eqnarray}
U_N=V_N/V_{N-1} \label{2.8}
\end{eqnarray}
and the diagonal part of $D$ on $U_N$ be defined as $D_N$
\begin{eqnarray}
D_N U_N = D U_N \cap U_N \label{2.9}
\end{eqnarray}
If the eigenvalues of $D_N$ are all different, the number of eigenvectors equals dim$U_N$. 
But if some eigenvalues coincide (this is true in the generic case!) the number of eigenvectors 
is smaller. Then the Hilbert space  on which the final selfadjoint Schr\"odinger operator is 
acting is not an $L^2$ -space. The missing eigenfunctions can be described. For more details 
see \cite{6}.

If we want completely integrable models we must make sure that a complete set of involutive 
differential operators exists. For this task Lie algebraic methods may be very helpful.

Given a differential operator (\ref{2.7}) one can characterize the vector $\vec{p}$ in (\ref{1}) 
by inequalities
\begin{eqnarray}
g_{[\vec{\alpha};a,b]} \neq 0 & \Rightarrow & \vec{p} \vec{\alpha} -p_a -p_b \leq 0 \label{2.10} \\
h_{[\vec{\beta};c]} \neq 0 & \Rightarrow & \vec{p} \vec{\beta} -p_c \leq 0 \label{2.11}
\end{eqnarray}
There should be enough equality signs in (\ref{2.10}),(\ref{2.11}) for a chosen $\vec{p}$ so that 
$D_N \neq 0$. It turns out that there exists a minimal $\vec{p}$-vector $\vec{p}_{\textrm{min}}$ 
so that the $V_N(\vec{p}_{\textrm{min}})$ spaces are maximal: For each $N,\vec{p}$ there is $N'$ 
so that 
\begin{eqnarray}
V_N(\vec{p}) \subset V_{N'}(\vec{p}_{\textrm{min}}) \label{2.12}
\end{eqnarray}
It is convenient to work only with this minimal $\vec{p}$-vector.
  
The first step in transforming $D$ into a Schr\"odinger operator is to write it symmetrically
\begin{equation}
D = - \sum_{a,b} \frac{\partial}{\partial z_a} g^{-1}_{ab} (z) \frac{\partial}{\partial z_b} +
\sum_a r_a(z) \frac{\partial}{\partial z_a} 
\label{8}
\end{equation}
where
\begin{equation}
g^{-1}_{ab} = \sum_{\vec{\alpha}} g_{[\vec{\alpha};a,b]} z^{[\vec{\alpha}]}
\label{9}
\end{equation}
We write $g^{-1}_{ab}$ because this is the inverse of a Riemann tensor. The Riemann tensor 
$g_{ab}$ is assumed to be curvature free. The task to make it so will not arise in this work. 
But we mention that we developed a minimal algorithm to solve this issue.

Following the notations of \cite{6} we "gauge" the polynomial eigenfunctions $\varphi$ of $D$ by
\begin{equation}
\psi(z) = e^{-\chi(z)} \varphi(z)
\label{10}
\end{equation}
so that
\begin{equation}
e^{-\chi} D e^{+\chi} = - \frac{1}{\sqrt{g}} \sum_{a,b} \frac{\partial}{\partial z_a} 
(\sqrt{g} g^{-1}_{ab}) \frac{\partial}{\partial z_b} + W(z)
\label{11}
\end{equation}
$(g = (\det g^{-1})^{-1})$. \\
This is possible if and only if
\begin{equation}
 \sum_b g^{-1}_{ab} (z) \frac{\partial}{\partial z_b} [ 2 \chi - \ln \sqrt{g} ] = r_a(z)
\label{12}
\end{equation}
which implies integrability constraints on the functions $\{r_a(z)\}$. If they are fulfilled 
we obtain a "prepotential"
\begin{equation}
\rho = \ln P
\label{13}
\end{equation}
so that
\begin{equation}
\rho = 2 \chi - \ln \sqrt{g}
\label{14}
\end{equation}

In most cases studied, we found solutions for $\rho$ as follows. Let
\begin{equation}
\det g^{-1}(z) = \prod^r_{i=1} P_i(z)
\label{15}
\end{equation}
where $\{P_i(z)\}$ are different real polynomials. Then
\begin{equation}
\rho(z) = \sum^r_{i=1} \gamma_i \ln P_i(z)
\label{16}
\end{equation}
with free parameters $\gamma_i$ solves the requirement that $\{r_a(z)\}$ (\ref{12}) belong 
to differential operators leaving each $V_N$ invariant. In particular
\begin{equation}
r_a^{(i)}(z) = \frac{1}{P_i(z)} \sum_b g^{-1}_{ab}(z) \frac{\partial P_i}{\partial z_b}
\label{17}
\end{equation}
are polynomials. Inserting (\ref{15}), (\ref{16}) 
in (\ref{14}) we obtain finally
 
\begin{equation}
\chi = \frac12 \sum^r_{i=1} (\gamma_i - \frac12) \ln P_i 
\label{18}
\end{equation}
We will later see that in the case of the models of Calogero type a term
\begin{equation}
\gamma_0 \ln P_0
\label{19}
\end{equation}
can be added to $\rho$, where
\begin{equation}
P_0(z) = e^{z_1}
\label{20}
\end{equation}
is not contained in $\det g^{-1}$ as a factor. This prepotential gives rise to the oscillator 
potential.

Finally we mention that $e^{-\chi}$ is the ground state wave function of the Schr\"odinger 
operator, as follows from (\ref{10}).

The expression \cite{6}, (6.17) for the potential $W(z)$ contains a term linear in $\chi$
\begin{equation}
- \sum_{a,b} \frac{\partial}{\partial z_a} \left( g^{-1}_{ab} \frac{\partial \chi}
{\partial z_b} \right) = 
 - \frac12 \sum^r_{i=1} (\gamma_i - \frac12) \sum_a \frac{\partial}{\partial z_a} r^{(i)}_a
\label{21}
\end{equation}
 Each divergence
\begin{equation}
\sum_a \frac{\partial}{\partial z_a} r^{(i)}_a (z) = C^{(i)}
\label{22}
\end{equation}
ought to be a constant. From now on we shall dismiss all constant terms in $W(z)$.

We can then write the potential as
\begin{equation}
W(z) = \sum_{i,j} \gamma_{ij} R_{ij}(z)
\label{23}
\end{equation}
\begin{equation}
R_{ij} = \sum_{a,b} g^{-1}_{ab} \frac{\partial \ln P_i}{\partial z_a} \frac
{\partial \ln P_j}{\partial z_b}
\label{24}
\end{equation}
\begin{equation}
\gamma_{ij} = \frac14 (\gamma_i\gamma_j - \frac14) \quad (i,j \not= 0).
\label{25}
\end{equation}
In the cases of this article
\begin{equation}
R_{ij} = {\rm const~if}\, i \not= j
\label{26}
\end{equation}
If we then set
\begin{equation}
\gamma_i = - \nu_i + \frac12 \quad (i \not= 0)
\label{27}
\end{equation}
we obtain
\begin{equation}
W(z) = \sum^r_{i=1} \gamma_{ii}R_{ii}(z)
\label{28}
\end{equation}
with
\begin{equation}
\gamma_{ii} = \frac14 \nu_i (\nu_i-1)
\label{29}
\end{equation}

As stated in the Introduction the variables $\{z_i\}$ appearing in this section are identified 
with Coxeter invariants formed from root space coordinates $\{x_n\}$ or $\{y_n\}$. These 
invariants are either polynomial or trigonometric. Finally we return from the invariant coordinates 
$\{z_i\}$ to the root space coordinates $\{x_n\}$ in the Schr\"odinger operator (\ref{11}). 
Each contribution
%
%
\begin{eqnarray}
R_{ii}=\frac{Q_{ii}}{P_{i}} \label{2.36}
\end{eqnarray}
admits a partial fraction decomposition due to the factorization of the prepotentials $P_i$ (Section 5). 
The label $i=1$ is always reserved to a "Vandermonde prepotential", i.e. 
%
%
%
\begin{eqnarray}
P_1 \sim \prod_{i<j} (x_i-x_j)^2 \quad \textrm{or} \quad  \prod_{i<j} (\sin(x_i-x_j))^2 \label{2.37}
\end{eqnarray}
or alike. 
\setcounter{equation}{0}
%
%
%
%
\setcounter{equation}{0}
\section{Translation invariant models}
\subsection{Relative coordinates}
The Laplacian for an Euclidean space $\bbbr_N$
\begin{equation}
\Delta = \sum^N_{i=1} \frac{\partial^2}{\partial x^2_i}
\label{3.1}
\end{equation}
is translation invariant. We introduce relative coordinates by
\begin{eqnarray}
y_i &=& x_i - \frac 1N X \\ \label{3.2}
X &=& \sum^N_{i=1} x_i
\label{3.3}
\end{eqnarray}
They separate the Laplacian such that
\begin{equation}
\Delta = N \frac{\partial^2}{\partial X^2} + \sum^N_{i=1} \frac{\partial^2}{\partial y^2_i} - 
\frac 1N \left( \sum^N_{i=1} \frac{\partial}{\partial y_i} \right)^2
\label{3.4}
\end{equation}
We use all $\{y_i\}^N_{i=1}$ as coordinates on the plane
\begin{equation}
\sum^N_{i=1} y_i = 0
\label{3.5}
\end{equation}
in order to maintain permutation symmetry.

\subsection{Elementary symmetric polynomials}
Elementary symmetric polynomials of $N$ variables $\{ q_i\}^N_{i=1}$ are defined by a
generating function
\begin{equation}
\sum^N_{n=0} p_n(q)t^n = \prod^N_{i=1} (1 + q_it) 
\label{3.6}
\end{equation}
They are invariant under the symmetric group $S_N$. For each $g \in S_N$ we have a sector 
(simplex) $E_g \subset \bbbr_N$
\begin{equation}
E_g = \{ q_{i_1} < q_{i_2} < \ldots < q_{i_N}; \quad i_n = g(n) \}
\label{3.7}
\end{equation}
so that
\begin{equation}
\bbbr_N = \bigcup_{g \in S_N} \bar{E}_g
\label{3.8}
\end{equation}
Inside $E_g$ we can use the $\{ p_n\}^N_{n=1}$ as coordinates since
\begin{equation}
{\cal M}_{ni} = \frac{\partial p_n}{\partial q_i}
\label{3.9}
\end{equation}
\begin{equation}
\det {\cal M} = (-1)^{[\frac N2]} V(q_1,q_2,...q_N)
\label{3.10}
\end{equation}
where $V$ is the Vandermonde determinant.

\subsection{The $A_{N-1}$ series}
The root system of $A_{N-1}$ and the corresponding Weyl group possess elementary symmetric 
polynomials as invariants. We express the Laplacian in each sector $E_g$ (\ref{3.7})
intersected with the plane (\ref{3.5}) in terms of these polynomials
\begin{equation}
\tau_n(y_1,...,y_N) = p_n(q)|_{q_i = y_i \, {\rm all} \, i}
\label{3.11}
\end{equation}
The dynamics will be bounded to such sectors by corresponding potential walls automatically.

Then (see [3]) it results
\begin{eqnarray}
\sum^N_{i=1} \frac{\partial^2}{\partial y^2_i} - \frac 1N \left( \sum^N_{i=1} \frac{\partial}
{\partial y_i} \right)^2  \nonumber \\
= \sum^N_{n,m=2} g^{-1}_{nm} \frac{\partial^2}{\partial \tau_n \partial \tau_m} +
\sum^N_{n=2} h_n \frac{\partial}{\partial \tau_n}
\label{3.12}
\end{eqnarray}
with
\begin{equation}
g^{-1}_{nm}(\tau) = \frac 1N (m-1)(N-n+1)\tau_n\tau_m - T_{n-1,m-1}(\tau)
\label{3.13}
\end{equation}
and
\begin{equation}
T_{nm}(\tau) = \sum_{l \ge 1} (2l+n-m)\tau_{n+l} \tau_{m-l}
\label{3.14}
\end{equation}
Here it is understood that
\begin{eqnarray}
\tau_0 &=& 1  \nonumber \\
\tau_1 &=& 0  \nonumber \\
\tau_n &=& 0 \; {\rm for} \; n < 0, \, n > N
\label{3.15}
\end{eqnarray}
In this case $\det g^{-1}$ is indecomposable as a polynomial, so we set
\begin{eqnarray}
P_0 &=& e^{\omega\tau_2} \\ \label{3.16}
P_1 &=& \det g^{-1} = C_N V(y_1,...,y_N)^2
\label{3.17}
\end{eqnarray}
The resulting vectors $\{r_a\}^N_2$ are
\begin{equation}
r^{(0)} = (-2\tau_2, - 3 \tau_3, ..., -N\tau_N) 
\label{3.18}
\end{equation}
\begin{equation}
r^{(1)} : \, \mbox{explicit formulas known only for} \, N \le 4
\label{3.19}
\end{equation}
and the potential is
\begin{equation}
\frac 12 W(x) = \frac 12 \omega^2 \sum^N_{i=1} x^2_i + g \sum_{1 \le i<j\le N} (x_i-x_j)^{-2}
\label{3.20}
\end{equation}

The corresponding Sutherland models are obtained as follows. We use as coordinates 
a system $\{ \sigma_n\}^N_{n=2}$ defined by (these differ from those in [3])
\begin{equation}
\sigma_0 = \prod^N_{i=1} \cos y_i
\label{3.21}
\end{equation}
and
\begin{equation}
\sigma_n = \sigma_0 \cdot p_n(q)|_{q_i = \tan y_i}
\label{3.22}
\end{equation}
The identity
\begin{eqnarray}
1 &=& \exp \left( i \sum^N_{j=1} y_j \right)  \nonumber \\
&=& \prod^N_{j=1} (\cos y_j + i \sin y_j)  \nonumber \\
&=& \sum^N_{n=0} i^n \sigma_n(y)
\label{3.23}
\end{eqnarray}
allows us to eliminate $\sigma_0$ and $\sigma_1$ in terms of the remaining $\{ \sigma_n\}^N
_{n=2}$ so that polynomials go into polynomials.

The Laplacian is expressed correspondingly as
\begin{eqnarray}
\sum^N_{i=1} \frac{\partial^2}{\partial y^2_1} - \frac 1N \left( \sum^N_{i=1} \frac{\partial}
{\partial y_i} \right)^2 = \nonumber \\
= \sum^N_{n,m=2} g^{-1}_{nm} \frac{\partial^2}{\partial \sigma_n \partial \sigma_m} +
\sum^N_{n=2} h_n \frac{\partial}{\partial \sigma_n}
\label{3.24}
\end{eqnarray}
\begin{eqnarray}
g^{-1}_{nm}(\sigma) &=& - T_{n+1,m+1}(\sigma) -T_{n+1,m-1}(\sigma)  \nonumber \\
& & - T_{n-1,m+1}(\sigma) - T_{n-1,m-1}(\sigma)  \nonumber \\
& & + \frac 1N [(m+1)\sigma_{m+1} + (m-1) \sigma_{m-1} ]  \nonumber \\
& & \times [(N-n-1) \sigma_{n+1} + (N-n+1)\sigma_{n-1}]
\label{3.25}
\end{eqnarray}
with $T_{nm}$ as in (\ref{3.14}).

Once again $\det g^{-1}$ is indecomposable, so we set
\begin{equation}
P_1 = \det g^{-1} = C^\prime_N \tilde{V} (y_1,...,y_N)^2
\label{3.26}
\end{equation}
where
\begin{equation}
\tilde{V} (y_1,...,y_N) = \prod_{i < j} \sin (y_i-y_j)
\label{3.27}
\end{equation}
has the symmetry of the Vandermonde determinant (translations and permutations). The vector
$r^{(1)}$ is known only up to $N = 4$. Finally we obtain as potential
\begin{equation}
\frac 12 W(x) = g \sum_{1 \le i<j \le N} \sin (x_i-x_j)^{-2}
\label{3.28}
\end{equation}
In each case $A_{N-1}$ the minimal $p$-vector is $(1,1,...,1) \in \bbbn^{N-1}$.

\subsection{The $G_2$ and $AG_3$ models}
The models $G_2$ and $AG_3$ belong also to the domain of translation invariant models
\cite{4}. For $G_2$ we start from $A_2$ and extend its Weyl group by a $\bbbz_2$ group
\[ y_i \to - y_i \]
As invariant variables we use \cite{4}
\begin{equation}
\lambda_2 = \tau_2
\label{3.29}
\end{equation}
\begin{equation}
\lambda_3 = \tau^2_3
\label{3.30}
\end{equation}
In these variables
\begin{eqnarray}
\sum^3_{i=1} \frac{\partial^2}{\partial y^2_i} - \frac 13 \left( \sum^3_{i=1} 
\frac{\partial}{\partial y_i} \right)^2 =  \nonumber \\
= \sum^3_{a,b=2} g^{-1}_{ab} \frac{\partial^2}{\partial \lambda_a \partial \lambda_b}
+ \sum^3_{a=2} h_a \frac{\partial}{\partial\lambda_a}
\label{3.31}
\end{eqnarray}
We find
\begin{equation}
g^{-1}(\lambda) = \left( \begin{array}{cc}
- 2 \lambda_2, & -6\lambda_3 \\
- 6 \lambda_3, & + \frac 83 \lambda^2_2 \lambda_3 \end{array} \right)
\label{3.32}
\end{equation}
so that
\begin{equation}
\det g^{-1} = - \frac 43 \lambda_3 (4 \lambda^3_2 + 27 \lambda_3)
\label{3.33}
\end{equation}
Thus as ansatz for the prepotentials we use 
\begin{eqnarray}
P_0 &=& e^{\omega\lambda_2} \\ \label{3.34}
P_1 &=& 4 \lambda^3_2 + 27 \lambda_3 \\ \label{3.35}
P_2 &=& \lambda_3
\label{3.36}
\end{eqnarray}
The $r$-vectors (justifying this ansatz) are
\begin{eqnarray}
r^{(0)} &=& (-2\lambda_2, - 6 \lambda_3)  \\ \label{3.37}
r^{(1)} &=& (-6,0) \\ \label{3.38}
r^{(2)} &=& (-6, + \frac 83 \lambda_2^2)
\label{3.39}
\end{eqnarray}
The minimal $\vec{p}$-vector is
\begin{equation}
\vec{p} = (1,2)
\label{3.40}
\end{equation}
The potential is
\begin{eqnarray}
\frac 12 W(x) &=& \frac 12 \omega^2 \sum^3_{i=1} x ^2_i \\ \nonumber
& & + g_1 \sum_{1 \le i<j \le 3} (x_i-x_j)^{-2} + g_2 \sum_{i<j,k \notin (i,j)}
(x_i+x_j-2x_k)^{-2}
\label{3.41}
\end{eqnarray}
with
\begin{eqnarray}
g_1 &=& \nu_1(\nu_1-1)  \nonumber \\
g_2 &=& 3\nu_2(\nu_2-1)
\label{3.42}
\end{eqnarray}
If
\begin{equation}
\nu_2 = 0 \; {\rm or} \; \nu_2 = 1
\label{3.43}
\end{equation}
we return to the $A_2$ model.

In the Sutherland case we use as variables
\begin{eqnarray}
\mu_2 &=& \sigma_2 \\ \label{3.44}
\mu_3 &=& \sigma^2_3
\label{3.45}
\end{eqnarray}
leading to the inverse Riemann tensor
\begin{equation}
g^{-1} = \left( \begin{array}{ll}
- 2 \mu_2 - 2 \mu^2_2 + \frac 23 \mu_3, & - \mu_3(6 + \frac{16}{3} \mu_2) \\ 
- \mu_3(6 + \frac{16}{3} \mu_2), & \frac 83 \mu^2_2 \mu_3 - 8 \mu^2_3 
\end{array} \right)
\label{3.46}
\end{equation}
Now $\det g^{-1}$ is decomposable with
\begin{equation}
\det g^{-1} = - \frac 43 \mu_3 P_1(\mu)
\label{3.47}
\end{equation}
and
\begin{equation}
P_1(\mu) = 4\mu^2_3 + \mu_3(8\mu^2_2 + 36\mu_2 + 27) + 4 \mu^3_2 (1 + \mu_2)
\label{3.48}
\end{equation}
\begin{equation}
P_2(\mu) = \mu_3
\label{3.49}
\end{equation}
The $r$-vectors are
\begin{eqnarray}
r^{(1)} &=& (-6-8\mu_2, - 16 \mu_3) \\ \label{3.50}
r^{(2)} &=& (-6 - \frac{16}{3} \mu_2, \frac 83 \mu^2_2 - 16 \mu_3)
\label{3.51}
\end{eqnarray}
The resulting potential is
\begin{eqnarray}
\frac 12 W(x) &=& g_1 \sum_{1 \le i < j \le 3} \sin (x_i-x_j)^{-2} \nonumber \\
& & + \frac 19 g_2 \sum_{i<j,k\notin (i,j)} \sin \frac13 (x_i+x_j-2x_k)^{-2}
\label{3.52}
\end{eqnarray}

In the case of the $A_2$ models the spaces $V_N$ decompose into even and odd 
subspaces in $\tau_3$ (or $\sigma_3$) which are left invariant separately under 
action of the Laplacian. In the case of the odd spaces we can factor $\tau_3 (\sigma_3)$ 
and leave an even space as well. In each case we obtain a polynomial space in the 
variables $\lambda_2, \lambda_3 = \tau^2_3 \, (\mu_2,\mu_3 = \sigma^2_3)$. Thus starting 
from such polynomial space and multiplying with $\tau_3^{\nu_2} \, (\sigma_3^{\nu_2})$ 
we obtain the $A_2$ model if $\nu_2=0$ or $\nu_2 =1$ but a new potential in all other cases.

It is plausible that a similar procedure works for $A_3$ but not for $A_{N-1}, \, N \ge 5$. 
In the latter models we have two or more odd variables $\tau_3, \tau_5,... (\sigma_3, 
\sigma_5,...)$ and there is no factorization of the odd invariant subspaces. Let us 
sketch the $A_3$ model whose extension leads to the $AG_3$ model \cite{8}.

In this case the variables are chosen as in (3.29), (3.30), (3.44), 
(3.45)
\begin{equation}
\lambda_2 = \tau_2, \; \lambda_3 = \tau^2_3, \; \lambda_4 = \tau_4
\label{3.53}
\end{equation}
The inverse Riemann tensor is
\begin{equation}
g^{-1} = \left( \begin{array}{ccc}
- 2 \lambda_2, & - 6 \lambda_3, & - 4 \lambda_4 \\
- 6 \lambda_3, & 4 \lambda_3(\lambda^2_2 - 4 \lambda_4), & \lambda_2 \lambda_3 \\
- 4 \lambda_4, & + \lambda_2 \lambda_3, & -2 \lambda_2 \lambda_4 + \frac34 \lambda_3 
\end{array} \right)
\label{3.54}
\end{equation}
The determinant is decomposable as 
\begin{equation}
\det g^{-1} = \lambda_3 P_1(\lambda)
\label{3.55}
\end{equation}
and the ansatz for the prepotentials is
\begin{eqnarray}
P_0(\lambda) &=& e^{\omega \lambda_2} \\ \label{3.56}
P_1(\lambda) &=& 27 \lambda^2_3 - 256 \lambda^3_4 + 128 \lambda^2_2 \lambda^2_4 \\ \nonumber
& & - 16 \lambda^4_2 \lambda_4 + 4 \lambda^3_2 \lambda_3 - 144 \lambda_2 \lambda_3 \lambda_4 \\ \label{3.57}
P_2(\lambda) &=& \lambda_3
\label{3.58}
\end{eqnarray}
The $r$-vectors come out as
\begin{eqnarray}
r^{(0)} &=& (-2 \lambda_2, - 6 \lambda_3, - 4 \lambda_4) \\ \label{3.59}
r^{(1)} &=& (-12, 0, - 2 \lambda_2) \\ \label{3.60}
r^{(2)} &=& (-6, 4(\lambda^2_2 - 4 \lambda_4), \lambda_2) \label{3.61}
\end{eqnarray}
The potential for this Calogero type model is
\begin{eqnarray}
\frac 12 W(x) &=& \frac12 \omega^2 \sum^4_{i=1} x^2_i \\ \nonumber
& & + g_1 \sum_{1 \le i < j \le 4} (x_i-x_j)^{-2} + g_2 \sum_{\rm 3 \, terms} 
(x_i+x_j-x_k-x_l)^{-2}
\label{3.62}
\end{eqnarray}
with
\begin{equation}
g_1 =\nu_1(\nu_1-1), \; g_2 = 2\nu_2(\nu_2-1)
\label{3.63}
\end{equation}
It was discovered first by Wolfes, \cite{7}.

The Sutherland model is obtained in the same fashion. With
\begin{equation}
\mu_2 = \sigma_2, \; \mu_3 = \sigma^2_3, \; \mu_4 = \sigma_4
\label{3.64}
\end{equation}
the inverse Riemann tensor is 
\begin{eqnarray}
g^{-1}_{22} &=& - 2 \mu_2 - 2\mu^2_2 - 8 \mu_4 + 2 \mu_3 + 8 \mu_2\mu_4 + 8 \mu^2_4 \\ \label{3.65}
g^{-1}_{23} &=& - 6 \mu_3 - 4\mu_2 \mu_3  \\ \label{3.66}
g^{-1}_{24} &=& - 4 \mu_4 - 6\mu_2 \mu_4 +  \mu_3 + 4 \mu^2_4  \\ \label{3.67}
g^{-1}_{33} &=& 4 \mu_3 [ - 4\mu_4 +  \mu^2_2 - 4 \mu_2\mu_4 + 4 \mu^2_4- 2 \mu_3] \\ \label{3.68}
g^{-1}_{34} &=&  \mu_2\mu_3 - 6 \mu_3\mu_4  \\ \label{3.69}
g^{-1}_{44} &=& - 2 \mu_2 \mu_4 + \frac 34 \mu_3   \label{3.70}
\end{eqnarray}
Its determinant decomposes
\begin{equation}
\det g^{-1} = - \mu_3 P_1(\mu)
\label{3.71}
\end{equation}
\begin{equation}
P_1(\mu) = 256 \mu^6_4 + 32 \; \mbox{further terms}
\label{3.72}
\end{equation}
\[ \mbox{ (equ. (A.2) from \cite{8})} \]
\begin{equation}
P_2(\mu) = \mu_3
\label{3.73}
\end{equation}
and the $r$-vectors are
\begin{equation}
r^{(1)} = (- 16\mu_2-12, - 24\mu_3, - 12 \mu_4-2\mu_2) 
\label{3.74}
\end{equation} 
\begin{equation}
r^{(2)} = (- 4 \mu_2-8, 16 \mu^2_4 - 16\mu_4\mu_2+4\mu^2_2 - 8 \mu_3 - 16 \mu_4, -6 \mu_4+\mu_2)
\label{3.75}
\end{equation}
The factorization of $\sigma_3$ which is necessary in this case is
\begin{equation}
\sigma_3 = - \prod_{1 \le i<j \le 3} \sin (y_i+y_j)
\label{3.76}
\end{equation}
implying
\begin{equation}
\frac{Q_{22}}{P_2} = 4 \sum_{1 \le i<j \le 3} (\sin (y_i+y_j))^{-2}
\label{3.77}
\end{equation}
This gives the potential
\begin{eqnarray}
\frac 12 W(x) &=& g_1 \sum_{1 \le i<j \le 4} (\sin (x_i-x_j))^{-2}  \nonumber \\
& & + \frac14 g_2 \sum_{3 \, {\rm cases}} (\sin \frac12 (x_i+x_j-x_k-x_l))^{-2}
\label{3.78}
\end{eqnarray}
The discussion of this $AG_3$ model is resumed in Section 5.

\setcounter{equation}{0}
\section{Translation non-invariant models}
\subsection{The $BC_N$ and $D_N$ models}
As we shall see there is only one series with two (Calogero) and three (Sutherland)
independent coupling constants. For any such model we use as Cartesian coordinates
$\{x_i\}^N_{i=1}$ and require permutation symmetry $S_N$ and reflection symmetry
$(\bbbz_2)^N$ $x_i \to - x_i$ for each $i$ separately. Then the natural coordinates
invariant under these group actions are \cite{5}
\begin{equation}
\lambda_n(x) = p_n(q)|_{q_i=x^2_i, \, {\rm all} \, i}
\label{4.1}
\end{equation}
There is a bilinear relation with the $\{ p_n(x)\}^N_{n=1}$
\begin{equation}
\lambda_n(x) = \sum^{2n}_{k=0} (-1)^{n-k} p_{2n-k} (x) p_k(x)
\label{4.2}
\end{equation}
The inverse Riemann tensor for the full Laplacian (\ref{3.1}) is then
\begin{equation}
g^{-1}_{nm}(\lambda) = 4 M_{nm}(\lambda)
\label{4.3}
\end{equation}
where we introduce the shorthand
\begin{equation}
M_{nm}(\lambda) = \sum_{l \ge 0} (2l+n-m+1) \lambda_{n+l} \lambda_{m-1-l}
\label{4.4}
\end{equation}
Its determinant factorizes
\begin{equation}
\det g^{-1} = (-1)^{\left[ \frac N2 \right]} 4^N \lambda_N P_1(\lambda)
\label{4.5}
\end{equation}
where
\begin{eqnarray}
P_1(\lambda) &=& N^N \lambda_N^{N-1} + ... \\ \nonumber
&=& D_N V(x_1^2,x_2^2,...x^2_N)^2
\label{4.6}
\end{eqnarray}
and
\begin{equation}
P_2(\lambda) = \lambda_N
\label{4.7}
\end{equation}
Both functions $P_1, P_2$ factorize in a trivial way. In the general case there is no
explicit expression for $r^{(1)}$ but
\begin{equation}
r_a^{(2)} = 4(N-a+1) \lambda_{a-1}
\label{4.8}
\end{equation}
If follows
\begin{equation}
R_{22} = 4 \frac{\lambda_{N-1}}{\lambda_N} = 4 \sum^4_{i=1} x^{-2}_i
\label{4.9}
\end{equation}
The resulting potential is, including an oscillator potential 
\begin{eqnarray}
\frac 12 W(x) &=& \frac 12 \omega^2 \sum^{N}_{i=1} x^2_i +  g_1 \sum_{1 \le i<j \le N} [(x_i-x_j)^{-2} + (x_i+x_j)^{-2}] \nonumber \\
& & + g_2 \sum^{N}_{i=1} x_i^{-2} \\ \label{4.10}
g_1 &=& \nu_1(\nu_1-1) \label{4.11} \\
g_2 &=&  \frac 12 \nu_2(\nu_2 -1)
\label{4.12}
\end{eqnarray}

In the Sutherland case we use coordinates
\begin{equation}
\mu_0 = \prod^N_{i=1} \cos^2 x_i
\label{4.13}
\end{equation}
\begin{eqnarray}
\mu_n(x) &=& \mu_0(x) p_n(q)|_{q_i = \tan^2x_i, \; {\rm all} \, i} \\ \nonumber
& & n \in \{ 1,2,...N\}
\label{4.14}
\end{eqnarray}
From the identity
\begin{eqnarray}
1 &=& \prod^N_{i=1} (\cos^2x_i + \sin^2x_i)  \nonumber \\
&=& \sum^N_{n=0} \mu_n(x)
\label{4.15}
\end{eqnarray}
we learn how to eliminate $\mu_0$ in facour of $\{\mu_n\}^N_{n_1}$ so that a polynomial
of $\{ \mu_n \}^N_{n=0}$ remains a polynomial.

In this case the inverse Riemannian is
\begin{eqnarray}
g^{-1}_{nm} &=& 4 \big\{ M_{n+1,m+1}(\mu) + M_{n,m}(\mu)  \nonumber \\
& & - M_{n,m+1}(\mu) - M_{n+1,m}(\mu) \big\}
\label{4.16}
\end{eqnarray}
and the determinant decomposes as
\begin{equation}
\det g^{-1} = 4^N (-1)^{\left[ \frac N2 \right]} \mu_0 \mu_N P_1(\mu)
\label{4.17}
\end{equation}
Now the factorization of $P_1(\mu)$ is
\begin{equation}
P_1(\mu) = D^\prime_N \prod_{1 \le i<j \le N} (\cos^2 x_i \sin^2x_j - \sin^2x_i \cos^2x_j)^2
\label{4.18}
\end{equation}
and we choose
\begin{eqnarray}
P_2(\mu) &=& \mu_N \\ \label{4.19}
P_3(\mu) &=& \mu_0 \label{4.20}
\end{eqnarray}
Again we have no general explicit expression for $r^{(1)}$ but
\begin{eqnarray}
r_a^{(2)} &=& 4[(N-a+1) \mu_{a-1} - (N-a) \mu_a ] \\ \label{4.21}
r_a^{(3)} &=& 4[(a+1) \mu_{a+1} - a \mu_a ] \label{4.22}
\end{eqnarray}
so that
\begin{eqnarray}
R_{22} &=& \frac{\mu_{N-1}}{\mu_N} = 4 \sum^N_{i=1} \cot^2 x_i \\ \label{4.23}
R_{33} &=& \frac{\mu_1}{\mu_0} = 4 \sum^N_{i=1} \tan^2 x_i \label{4.24}
\end{eqnarray}
Thus we end up with a potential
\begin{eqnarray}
 \frac 12 W(x) &=& g_1 \sum_{1\le i<j \le N} [(\sin (x_i-x_j))^{-2} + (\sin (x_i+x_j))^{-2}  ]  \nonumber \\
& & + g_2 \sum^N_{i=1} (\sin x_i)^{-2}  \nonumber \\
& & + g_3 \sum^N_{i=1} (\cos x_i)^{-2}
\label{4.25}
\end{eqnarray}
where $g_{1,2}$ are as in (\ref{4.11}),(\ref{4.12}) and 
\begin{eqnarray}
g_3=\frac 12 \nu_3 (\nu_3-1) \label{4.25a}
\end{eqnarray}
An alternative form of the potential is obtained from
\begin{equation}
\frac{g_2}{\sin^2 x} + \frac{g_3}{\cos^2x} = \frac{g_2-g_3}{\sin^2x} + 
\frac{4g_3}{\sin^2 2x}
\label{4.26}
\end{equation}
If we set $g_2=g_3$ or $g_3=0$ we obtain different samples of the $BC_N$ or $D_N$ series.
We mention finally that the minimal $p$-vector is in all cases
\begin{equation}
\vec{p} = (1,1,...1) \in \bbbn^N
\label{4.27}
\end{equation}

\subsection{The $F_4$ model}
The $F_4$ model belongs also to the translation noninvariant class. The Weyl group of
$F_4$ possesses four basic polynomial invariants
\begin{equation}
I_1(x), \, I_3(x), \, I_4(x), I_6(x)
\label{4.28}
\end{equation}
($I_n$ of degree $2n$) which can be expressed as polynomials in the $\{\lambda_n \}^4_{n=1}$ as
follows
\begin{eqnarray}
I_1 &=& \lambda_1 \\ \label{4.29}
I_3 &=& \lambda_3 - \frac16 \lambda_1 \lambda_2 \\ \label{4.30}
I_4 &=& \lambda_4 - \frac14 \lambda_1 \lambda_3 + \frac{1}{12} \lambda^2_2 \\ \label{4.31}
I_6 &=& \lambda_4 \lambda_2 - \frac{1}{36} \lambda^3_2 + \frac{1}{24} \lambda^2_2 \lambda^2_1
- \frac{1}{64} \lambda_2 \lambda^4_1
\label{4.32}
\end{eqnarray}
In these coordinates the inverse Riemannian can be given as
\begin{eqnarray}
g^{-1}_{1m} &=& 4 m I_m \\ \label{4.33}
g^{-1}_{33} &=& \frac{20}{3} I_4 I_1 - \frac23 I_3 I^2_1 \\ \label{4.34}
g^{-1}_{34} &=& 8I_6 - 3 I^2_3 - \frac{13}{3} I_4I^2_1 - \frac{3}{4} I_3I^3_1
\\ \label{4.35}
g^{-1}_{36} &=& 16I_4^2 + I_6I^2_1 + 14 I_4I_3I_1 + \frac 52 I_3^2I^2_1
- \frac{1}{4} I_4I^4_1 - \frac{5}{32} I_3I^5_1
\\ \label{4.36}
g^{-1}_{44} &=& -4I_4I_3 - 2 I_6I_1 + \frac{3}{4} I_4I^3_1
+ \frac{3}{4} I^2_3I_1 + \frac{3}{16} I_3I^4_1
\\ \label{4.37}
g^{-1}_{46} &=& 8I^2_4I_1 + 2 I_4I_3I^2_1 - \frac{1}{8} I_4I^5_1 \\ \label{4.38}
g^{-1}_{66} &=& 30 I_6I_4I_1 + \frac{21}{2} I_6I_3I^2_1 - \frac{3}{32} I_6I^5_1 
+ 12 I^2_4I_3 + 6 I_4I^2_3I_1 \nonumber \\
& &  - \frac{3}{8} I_4I_3I^4_1 + \frac{3}{4} I_3^3I^2_1 
+ \frac{3}{1024} I_3I^8_1 - \frac{3}{32} I^2_3I^5_1 \label{4.39}
\end{eqnarray}
The determinant decomposes into two factors 
\begin{equation}
\det g^{-1} = \frac{1}{3072} P_1(I) P_2(I)
\label{4.40}
\end{equation}
where $P_1(I)$ is connected with the Vandermonde determinant squared as usual
\begin{eqnarray}
P_1(I) &=& - 4096 I^3_4 + 432 I^4_3 + 3072 I^2_6 - 2304 I_6I_4I^2_1  \nonumber \\
& & - 576 I_6 I_3 I^3_1 + 864 I_4I^2_3I^2_1 + 216 I_4 I_3 I^5_1  \nonumber \\
& & + 432 I^2_4 I^4_1 + 27 I^2_3 I^6_1 - 2304 I_6 I^2_3 + 216 I^3_3 I^3_1
\label{4.41}
\end{eqnarray}
or in factorized form
\begin{equation}
P_1(I) = - 16 \prod_{1 \le i<j \le 4} (x^2_i-x^2_j)^2
\label{4.42}
\end{equation}
and $P_2(I)$
\begin{eqnarray}
P_2(I) &=&  36864 I^2_6 - 18432 I_6I_4I^2_1 - 4608 I_6I_3I^3_1 + 32 I_6I^6_1  \nonumber \\
& & - 49152 I^3_4 - 36864  I^2_4I_3I_1 + 1536 I^2_4 I^4_1 \nonumber \\
& & + 768 I_4I_3I^5_1 - 12 I_4I^8_1 - 9216 I_4I^2_3I^2_1  \nonumber \\
& & - 768 I^3_3 I^3_1 + 96 I^2_3I^6_1 - 3I_3I^9_1
\label{4.43}
\end{eqnarray}
which factorizes as
\begin{eqnarray}
P_2(I) &=& - 12\lambda_4 (64 \lambda_4-16\lambda^2_2 + 8\lambda_2\lambda^2_1 - \lambda^4_1)^2 
\nonumber \\
&=& - 12 x^2_1x^2_2x^2_3x^2_4 \prod_{\nu_2,\nu_3\nu_4 \in \{1,0\}} (x_1 - \sum^4_{i=2}
(-1)^{\nu_i}x_i)^2 
\label{4.44}
\end{eqnarray}
The $r$-vectors are
\begin{eqnarray}
r^{(1)} = (48, - 2 I^2_1, 0, 36I_4I_1 + 12I_3I^2_1 - \frac{3}{16} I^5_1) \\ \label{4.45}
r^{(2)} = (48, - 4 I^2_1, - 12I_3, 24 I_4I_1 + 6 I^2_1I_3 - \frac{3}{8} I^5_1)
\label{4.46}
\end{eqnarray}
The potential resulting is
\begin{eqnarray}
\frac 12 W(x) &=& \frac12 \omega^2 \sum_{1 \leq i \leq 4} x^2_i +  g_1 \sum_{1 \le i < j \le 4} [(x_i-x_j)^{-2} + (x_i+x_j)^{-2}]  \nonumber \\
& & +  g_2 \big\{ \sum_{\begin{array}{c} {\scriptstyle
\nu_2, \nu_3, \nu_4} \\ {\scriptstyle \in \{ +1,0\}} \end{array}}
4 \left( x_1 - \sum^4_{i=2} \nu_i x_i \right)^{-2} + \sum^4_{i=1} x^{-2}_i \big\}
\label{4.47}
\end{eqnarray}
where $g_{1,2}$ are as in (\ref{4.11}),(\ref{4.12}).
The minimal $p$-vector is
\begin{equation}
\vec{p} = (1,2,3,5)
\label{4.49}
\end{equation}

\setcounter{equation}{0}
\section{Coxeter groups, orbits and prepotentials}
The prepotentials used in the empirical constructions of sections 3 and 4 necessitate a mathematical 
interpretation. Let $W$ be a Coxeter group generated by the reflections 
\begin{equation}
\{s_{\alpha}\}
\end{equation}
where $\alpha$ are roots running over a set 
\begin{equation}
\Phi = \{\alpha\}_1^M
\end{equation}
The roots span an Euclidian space $V$. In this space the reflections $\{s_{\alpha}\}$ act by 
\begin{eqnarray}
x \in V: s_{\alpha} x = x - 2 \frac{(\alpha,x)}{(\alpha,\alpha)} \alpha
\end{eqnarray}
If the Coxeter group $W$ is "crystallographic", it is a Weyl group (for more details see \cite{9}). 

We denote a set of basic polynomial invariants of $W$ by 
\begin{eqnarray}
\{z_1(x), \ldots, z_n(x)\}, \quad  n=\dim V
\end{eqnarray}
Invariance means 
\begin{eqnarray}
z_i(w^{-1}x) & = & z_i(x) \nonumber \\
             & = & wz_i(x)
\end{eqnarray}
for all $w\in W$. The Jacobian for the transition $\{x_j\} \rightarrow \{z_i\}$ 
\begin{eqnarray}
J = \det \left \{ \frac{\partial z_i}{\partial x_j} \right \}
\end{eqnarray}
can be factorized as follows (\cite{9}, Proposition 3.13).

Each reflection $s_{\alpha}$ leaves a hyperplane $H_{\alpha}$ in $V$ pointwise fixed, let 
$H_{\alpha}$ be given by a linear function $l_{\alpha}$ 
\begin{equation}
l_{\alpha}(x) = 0 \label{5.7}
\end{equation}
Then due to the proposition 
\begin{equation}
J = C \prod_{\alpha \in \Phi^{+}} l_{\alpha}(x)
\end{equation}
with $\Phi^{+}$ the set of positive roots. The proof of this proposition is rather elementary. 

For any inverse Riemann tensor $\{ g^{-1} \}$ of Sections 3 and 4 we obtain this way 
\begin{equation}
\det{g^{-1}_{ab}} = C^2 \prod_{\alpha \in \Phi^{+}} l_{\alpha}(x)^2 \label{5.9}
\end{equation}
If $\Phi$ decomposes into orbits under $W$
\begin{equation}
\Phi = \bigcup_i \Phi_i
\end{equation}
then
\begin{equation}
P_i = \prod_{\alpha \in \Phi^{+}_i} l_{\alpha}(x)^2 \label{5.11}
\end{equation}
is an invariant polynomial under action of $W$ and therefore a polynomial in the basic invariants
\begin{equation}
P_i = P_i(z_1, \ldots , z_n)
\end{equation}
These polynomials are the prepotentials constructed in Sections 3 and 4. The factorization of 
these prepotentials as quoted at the end of Section 2 (eqns. (\ref{2.36}),(\ref{2.37})) and 
used throughout in Sections 3 and 4 is based on (\ref{5.11}).

We emphasize that our empirical results of Sections 3 and 4 indicate the validity of further 
mathematical propositions which could not be traced in the literature:
\begin{enumerate}
\item an analogous factorization theorem for the trigonometric invariants;
\item the polynomial properties ("integrability") of the functions $r^{(i)}(z)$ (\ref{17}).
\end{enumerate}

Now we return to the $AG_3$ model of Section 3. We identify the roots involved in a model using (\ref{5.7}),(\ref{5.9})
\begin{eqnarray}
l_{\alpha}(x) & = & (\alpha^{\vee},x) \nonumber \\
(\alpha^{\vee} & = & \frac{2 \alpha}{(\alpha,\alpha)}, \textrm{ the "dual" of } \alpha ) \label{5.13}
\end{eqnarray}
and the Sutherland version whose potential is 
\begin{eqnarray}
\frac12 W(x) = \sum_{\begin{array}{c} \textrm{\footnotesize{orbits }} i \end{array}} g_i \sum_{\alpha \in \Phi^{+}_i} [\sin l_{\alpha}(x)]^{-2} \label{5.14} 
\end{eqnarray}
Thus the simple roots of $A_3$ 
\begin{eqnarray}
\alpha_1 & = & e_1 - e_2 \nonumber \\
\alpha_2 & = & e_2 - e_3 \label{5.15} \\
\alpha_3 & = & e_3 - e_4 \nonumber
\end{eqnarray}
are completed by a fourth root in $AG_3$ 
\begin{eqnarray}
\alpha_4 & = & e_3 + e_4 - e_1 - e_2 \label{5.16}
\end{eqnarray}
The corresponding Coxeter-diagram is shown in Fig. 1.
%
%
%
%
\begin{figure}
\begin{center}
\epsfig{file=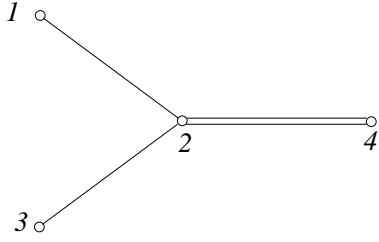,height=3cm,angle=0}
\end{center} 
\label{fig:coc_b3_hut}
\caption{Coxeter diagram of $\hat{B}_3$} 
\end{figure}
It belongs to the affine Coxeter group $\hat{B}_3$ (\cite{9}, Figure 1 in Section 2.4).

The coordinates of the $\hat{B}_3$ root space with respect to the standard basis $\{f_i\}_{i=1}^3$ 
are denoted $\{\xi_i\}_{i=1}^3$, those of $AG_3$ with respect to the standard basis $\{e_i\}_{i=1}^4$ by 
$\{x_i\}_{i=1}^4$ as before. The simple roots of $B_3$ are 
\begin{eqnarray}
\beta_1 = f_1 - f_2, \quad \beta_1 = f_2 - f_3, \quad \beta_3 = f_3 \label{5.17}
\end{eqnarray}
and $\hat{B}_3$ is obtained by adjoining 
\begin{eqnarray}
\beta_4 = - f_1 - f_2 \label{5.18}
\end{eqnarray}
It follows that 
\begin{eqnarray}
s_4 \left( \begin{array}{c} \xi_1 \\ \xi_2 \\ \xi_3 \end{array} \right) = \left( \begin{array}{c} -\xi_2 \\ -\xi_1 \\ \xi_3 \end{array} \right) \label{5.19}
\end{eqnarray}
leaves the Coxeter invariants of $B_3$ 
\begin{eqnarray}
\lambda_1(\xi) & = & \sum_{1 \leq i \leq 3} \xi_i^2 \label{5.20} \\
\lambda_2(\xi) & = & \sum_{1 \leq i < j \leq 3} \xi_i^2 \xi_j^2 \label{5.21} \\
\lambda_3(\xi) & = & \xi_1^2 \xi_2^2 \xi_3^2 \label{5.22} 
\end{eqnarray}
invariant, too. This suggests the equivalence of the $AG_3$ and the $B_3$ models. 

An explicit identification of the simple roots 
\begin{eqnarray}
f_1 & = & \frac12 (e_1 - e_2 - e_3 +e_4) \label{5.23} \\
f_2 & = & \frac12 (- e_1 + e_2 - e_3 +e_4) \label{5.24} \\
f_3 & = & \frac12 (- e_1 - e_2 + e_3 +e_4) \label{5.25} 
\end{eqnarray}
gives ($i,j \in \{1,2,3\}$) 
\begin{eqnarray}
x_i-x_j & = & \xi_i - \xi_j \label{5.26} \\
x_4 - x_j & = & \sum_{i (\neq j)} \xi_i \label{5.27}
\end{eqnarray}
It follows 
\begin{eqnarray}
& & g_1 \sum_{1 \le i<j \le 4} [\sin (x_i-x_j)]^{-2} + \frac14 g_2 \sum_{3 \, {\rm cases}} [\sin \frac12 (x_i+x_j-x_k-x_l)]^{-2} \nonumber \\
&=& g_1 \sum_{1\le i<j \le 3} \{ [\sin (\xi_i-\xi_j)]^{-2} + [\sin (\xi_i+\xi_j)]^{-2} \} + \frac14 g_2 \sum^3_{i=1} [\sin \xi_i]^{-2} \nonumber \\  \label{5.28}
\end{eqnarray}
Moreover the rational invariants (\ref{3.64}) can be identified with the 
invariants (\ref{5.20})--(\ref{5.22})
\begin{eqnarray}
\mu_2(x) & = & - \frac12 \lambda_1(\xi) \label{5.29} \\
\mu_3(x) & = & + \frac14 \lambda_3(\xi) \label{5.30} \\
\mu_4(x) & = & - \frac14 \lambda_2(\xi) +\frac{1}{16} \lambda_1(\xi)^2 \label{5.31} 
\end{eqnarray}
This establishes the equivalence between the two models. 

Our method involves a reduction of the affine Coxeter group $\hat{B}_3$ to the Coxeter 
group $B_3$ having the same invariants. It may therefore be of interest that the construction 
performed in \cite{4} is analogous (see Fig. 2).
%
%
%
%
\begin{figure}
\begin{center}
\epsfig{file=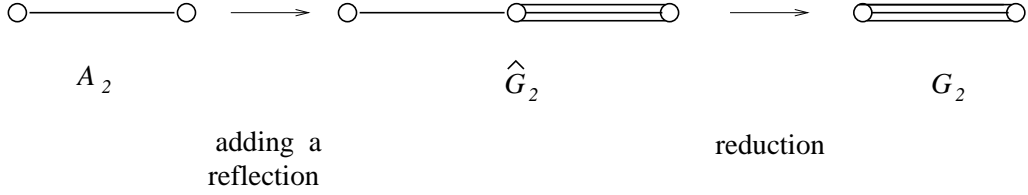,height=2.5cm,angle=0}
\end{center} 
\label{fig:coxag2_a}
\caption{Extending the Coxeter diagram of $A_2$ to $\hat{G}_2$ and reduction to $G_2$} 
\end{figure} 

\newpage
\begin{thebibliography}{99}
\bibitem{1} M.A: Olshanetsky and A.M. Perelomov, Lett. Math. Phys. 2 (1977) 7-13, and Phys. 
Reports 94 (1983) 313.
\bibitem{2} A.V. Turbiner, "Lie algebras and linear operators with invariant subspace", in 
"Lie algebras, cohomologies and new findings in quantum mechanics" (N. Kamran and P.J. 
Olver eds.) AMS, Vol 160, pp. 263-310, 1994; \\
"Lie-algebras and quasi-exactly-solvable differential equations", Vol 3: New Trends in 
Theoretical Developments and Computational Methods, Chapter 12, CRC Press (N. Ibragimov ed.) 
pp. 331-366, 1995 (hep-th 9409068).
\bibitem{3} W. R\"uhl and A. Turbiner, Mod. Phys. Letters A10 (1995) 2213 \\ (hep-th 9506105).
\bibitem{4} M. Rosenbaum, A. Turbiner, A. Capella, Solvability of the $G_2$ Integrable System; 
(sol-int 9707005).
\bibitem{5} L. Brink, A. Turbiner, N. Wyllard, J. Math. Phys. 39 (1998) 1285.
\bibitem{6} O. Haschke, W. R\"uhl, Exactly solvable dynamical systems in the neighborhood 
of the Calogero model, to appear (hep-th 9803169); in eq.
(7.14) $a_7$ should read $a_7-3$.
\bibitem{7} J. Wolfes, Ann. Phys. (N.Y.) 85 (1974) 454. 
\bibitem{8} O. Haschke, W. R\"uhl, Construction of exactly solvable quantum models of 
Calogero and Sutherland type with translation invariant four-particle interactions, (hep-th 9807194) 
\bibitem{9} J.E. Humphreys, Reflection groups and Coxeter groups, Cambridge University Press 1990
\end{thebibliography}
\end{document}